\documentclass[trackchanges,manuscript]{aastex63}

\usepackage{ulem}
\usepackage{amsmath}
\usepackage{epstopdf}

\def\hho  {H$_2$O}

\def\kms  {km~s$^{-1}$}
\def\masy {mas~y$^{-1}$}

\def\deg  {\ifmmode {^\circ}\else {$^\circ$}\fi}

\def\etal {et al.~}

\def\p    {\phantom{0}}
\def\d    {\ifmmode {{\rlap{.}}^\circ}\else {${\rlap{.}}^\circ$}\fi}
\def\s    {\ifmmode {{\rlap{.}}^s}\else {${\rlap{.}}^s$}\fi}
\def\as   {\ifmmode {{\rlap{.}}^{''}}\else {${\rlap{.}}^{''}$}\fi}

\def\chisqpdf {\ifmmode {\chi^2_{\rm pdf}}\else {$\chi^2_{\rm pdf}$}\fi}
\def\chisq    {\ifmmode {\chi^2}\else {$\chi^2$}\fi}
\def\pa    {\ifmmode {\psi} \else {$\psi$}\fi}

\newbox\grsign \setbox\grsign=\hbox{$>$} \newdimen\grdimen \grdimen=\ht\grsign
\newbox\laxbox \newbox\gaxbox
\setbox\gaxbox=\hbox{\raise.5ex\hbox{$>$}\llap
{\lower.5ex\hbox{$\sim$}}}\ht1=\grdimen\dp1=0pt
\setbox\laxbox=\hbox{\raise.5ex\hbox{$<$}\llap
{\lower.5ex\hbox{$\sim$}}}\ht2=\grdimen\dp2=0pt

\def\lax{\mathrel{\copy\laxbox}}

\def\vlsr  {\ifmmode {v_{\rm LSR}}\else {$v_{\rm LSR}$}\fi}
\def\Vlsr {\ifmmode {V_{\rm LSR}} \else {$V_{\rm LSR}$} \fi}
\def\Vlsra {\ifmmode {{V_{\rm LSR}}^{a}} \else {$V_{\rm LSR}^{\ast}$} \fi}
\def\vhelio{\ifmmode {v_{Helio}}\else {$v_{Helio}$}\fi}

\def\Ge {G020.08$-$0.13}
\def\Gd {G019.60$-$0.23}
\def\Gc {G020.77$-$0.05}
\def\Gb {G026.50$+$0.28}

\def\To    {\ifmmode{\Theta_0}\else{$\Theta_0$}\fi}
\def\Ro    {\ifmmode{R_0}\else{$R_0$}\fi}
\def\Ts    {\ifmmode{\Theta_s}\else{$\Theta_s$}\fi}
\def\Tdot  {\ifmmode{d\Theta\over dR}\else{$d\Theta\over dR$}\fi}

\def\ura   {\ifmmode {\mu_\alpha}\else {$\mu_\alpha$}\fi}
\def\udec  {\ifmmode {\mu_\delta}\else {$\mu_\delta$}\fi}

\def\ul    {\ifmmode {\mu_l}\else {$\mu_l$}\fi}
\def\ub    {\ifmmode {\mu_b}\else {$\mu_b$}\fi}

\def\uml   {\ifmmode {v_{gr}}\else {$v_{gr}$}\fi}
\def\umb   {\ifmmode {v_b}\else {$v_b$}\fi}
\def\vsrad {\ifmmode {v_{rad}}\else {$v_{rad}$}\fi}

\def\upl   {\ifmmode {v^p_{gr}}\else {$v^p_{gr}$}\fi}
\def\upb   {\ifmmode {v^p_b}\else {$v^p_b$}\fi}
\def\vprad {\ifmmode {v^p_{rad}}\else {$v^p_{rad}$}\fi}

\def\Vo    {\ifmmode {V^{Std}_\odot}\else {$V^{Std}_\odot$}\fi}
\def\Uo    {\ifmmode {U^{Std}_\odot}\else {$U^{Std}_\odot$}\fi}
\def\Wo    {\ifmmode {W^{Std}_\odot}\else {$W^{Std}_\odot$}\fi}

\def\V     {\ifmmode {V_\odot}\else {$V_\odot$}\fi}
\def\U     {\ifmmode {U_\odot}\else {$U_\odot$}\fi}
\def\W     {\ifmmode {W_\odot}\else {$W_\odot$}\fi}

\def\Vs    {\ifmmode {V_s}\else {$V_s$}\fi}
\def\Us    {\ifmmode {U_s}\else {$U_s$}\fi}
\def\Ws    {\ifmmode {W_s}\else {$W_s$}\fi}

\shortauthors{Xu et al.}

\begin{document}

\title{Trigonometric Parallaxes of Four Star-forming Regions in the Distant Inner Galaxy}
\author{Y. Xu}
\affiliation{Purple Mountain Observatory, Chinese Academy of Sciences,
Nanjing 210033, China; xuye@pmo.ac.cn}
\author{S. B. Bian}
\affiliation{Purple Mountain Observatory, Chinese Academy of Sciences,
Nanjing 210033, China; xuye@pmo.ac.cn}
\affiliation{School of Astronomy and Space Science, University of
Science and Technology of China, Hefei 230026, China}
\author{M. J. Reid}
\affiliation{Center for Astrophysics~$\vert$~Harvard $\&$ Smithsonian,
60 Garden Street, Cambridge, MA 02138, USA}
\author{J. J. Li}
\affiliation{Purple Mountain Observatory, Chinese Academy of Sciences,
Nanjing 210033, China; xuye@pmo.ac.cn}
\author{K. M. Menten}
\affiliation{Max-Planck-Institut f$\ddot{u}$r Radioastronomie, Auf
dem H{\" u}gel 69, 53121 Bonn, Germany}
\author{T. M. Dame}
\affiliation{Center for Astrophysics~$\vert$~Harvard $\&$ Smithsonian,
60 Garden Street, Cambridge, MA 02138, USA}
\author{B. Zhang}
\affiliation{Shanghai Astronomical Observatory, Chinese Academy of
Sciences, Shanghai 200030, China}
\author{A. Brunthaler}
\affiliation{Max-Planck-Institut f$\ddot{u}$r Radioastronomie, Auf
dem H{\" u}gel 69, 53121 Bonn, Germany}
\author{Y. W. Wu}
\affiliation{National Time Service Center, Chinese Academy of Sciences,
Xi'an 710600, China}
\author{L. Moscadelli}
\affiliation{INAF-Osservatorio Astrofisico di Arcetri, Largo E. Fermi
5, 50125 Firenze, Italy}
\author{G. Wu}
\affiliation{Xinjiang Astronomical Observatory, Chinese Academy of
Sciences, Urumqi 830011, China}
\author{X. W. Zheng}
\affiliation{Nanjing University, Nanjing 20093, China}

\begin{abstract}

  We have measured trigonometric parallaxes for four \hho\ masers
  associated with distant massive young stars in the inner regions
  of the Galaxy using the VLBA as part of the BeSSeL Survey.
  \Gb\ is located at the near end of the Galactic bar, \edit3{perhaps at the origin} of the Norma spiral arm.  \Gc\ is in
  the Galactic Center region and is likely associated with a far-side extension
  of the Scutum arm. \Gd\ and \Ge\ are \edit3{likely associated and lie} well past the Galactic Center.
  These sources appear to be in the 
  Sagittarius spiral arm, \edit3{but an association with the Perseus arm cannot be ruled out.}

\end{abstract}

\section{Introduction}
Mapping the Milky Way's spiral structure is a challenging enterprise,
since distances can exceed 10 kpc and dust in the plane obscures
optical light.  The use of maser parallax measurements with Very Long Baseline Interferometry (VLBI) \citep{xrzm06,Hachi:06} has allowed astronomers to determine
far more precise distances. Recently, \citet{VERA:2020} published
a compilation of astrometry results.
Together with the National Radio Astronomy Observatory (NRAO)
Very Long Baseline Array's (VLBA) key science project, the
Bar and Spiral Structure Legacy (BeSSeL) Survey summarized in \citet{Reid:19},
a total of more than 200 parallaxes and proper motions of maser sources
have been measured.  A significant portion of the spiral arms of the
Milky Way have now been characterized \citep[e.g.,][]{Choi:14,Wu:14,Wu:19,Sato:14,Reid:19},
especially within 5 kpc of the Sun \citep{Xu13,Xu16,Hodges:16}.
However, for very distant regions of the Milky Way, especially towards
the Galactic Center, a paucity of parallax measurements still limits our
understanding of the \edit3{Galactic} spiral structure. In order to improve
upon this, as part of the BeSSeL Survey, we present trigonometric
parallaxes of four distant star-forming regions toward the inner Galaxy.

\section{Observations and Analysis}
We observed 22-GHz \hho\ masers towards four star-forming regions over 16 epochs
with the VLBA, under program BR210; each epoch consisted of a 7-hr track.
We scheduled observations near the peaks of the sinusoidal parallax signature
in right ascension, since the amplitude of the declination signature is considerably
smaller. Calibration procedures followed our well documented BeSSeL Survey
\citep{Reid:09}.

The parallax and proper motion of each source were fitted simultaneously
using the obtained data.  We added ``error floors'' in quadrature to the
formal position uncertainties to achieve post-fit residuals with $\chi^2$
per degree of freedom near unity for both coordinates.
Since uncompensated tropospheric delay differences between the masers and
the background quasars usually dominate the systematics, and these delay
differences are essentially the same for all maser spots in a source,
we conservatively inflated formal parallax errors by the square-root of the
number of maser spots used.
Table~\ref{table:parallax} summarizes parallaxes and proper motions results.
\edit3{Additional details on} observations and analysis are presented in the \nameref{sec:app}.

\begin{deluxetable*}{lrrrrr}
	\tablecolumns{6} \tablewidth{0pc} \tablecaption{Parallaxes and Proper
		Motions. \label{table:parallax}}
	\tablehead{ \colhead{Source} & \colhead{$\pi$} & \colhead{D} & \colhead{$\mu_x$}
		& \colhead{$\mu_y$} & \colhead{\Vlsra}\\
		\colhead{} & \colhead{(mas)} & \colhead{(kpc)}  & \colhead{(\masy)}
		& \colhead{(\masy)} & \colhead{(\kms)}} \startdata
	\Gd & $0.076\pm0.011$ &  $13.2^{+ 2.2}_{- 1.7}$ & $-3.11\pm0.16$
	&$-6.36\pm0.17$ & $41^{a}\pm 3 $\\
	\Ge  & $0.066\pm0.010$ & $15.2^{+ 2.7}_{- 2.0}$ & $-3.14\pm0.14$
	& $-6.44\pm0.16$ &  $ 41^{b}\pm 3 $ \\
	\Gc & $0.124\pm0.013$ & $ 8.1^{+ 0.9}_{- 0.8}$& $-3.27\pm0.26$
	& $-6.55\pm0.27$ &  $57^{a}\pm 3$ \\
	\Gb  & $0.159\pm0.012$ & $ 6.3^{+ 0.5}_{- 0.4}$ & $-2.51\pm0.34$
	& $-6.04\pm0.34$ &  $ 104^{c}\pm 3 $ \\
	\enddata
	\tablecomments {Columns 1 gives the Galactic source name/coordinates.
		Columns 2 through 5 give the parallax, parallax distance and proper
		motion in the
		eastward ($\mu_x$ = $\mu_\alpha\cos\delta$) and northward directions
		($\mu_y$ =$\mu_\delta$ ). Column 6 lists the local standard of rest velocity
		from molecular line emission.
		\Vlsr references: (a) \citet{Wienen:12}, NH$_{3}$ $(J, K)$ = (1, 1);
		(b) \citet{Dunham:11}, NH$_{3}$ $(J, K)$ = (1, 1); (c) \citet{Yang:17}, CH$_{3}$OH.}
\end{deluxetable*}

\edit3{Since water masers in star forming regions form in outflows with expansion velocities typically 10s of \kms, assigning a \Vlsr to the central star which excites the masers can be uncertain at the $\pm$10 \kms\ level.   Therefore, we instead adopt the \Vlsr measured with dense molecular tracers (e.g., NH$_{3}$ and CH$_{3}$OH) for the associated molecular cloud.}

\edit3{Our distances to \Gd\ and \Ge\ resolve a long-standing controversy
as to whether the sources are at their near kinematic distance
\citep[$\sim$4 kpc,][]{GG:76,Petriella:13} or the far
\citep[$\sim$13 kpc,][]{Kolpak:03,Rana:18}. Their parallax distances
of $13.2^{+ 2.2}_{- 1.7}$ kpc and $15.2^{+ 2.7}_{- 2.0}$ kpc
strongly favor the far distance.
The kinematic distance of \Gc\ is $\sim$11.6 kpc \citep{Urquhart:18}
based on the HI self-absorption technique, but its parallax distance
of $8.1^{+ 0.9}_{- 0.8}$ kpc is much less, suggesting its luminosity
is just half its previous estimate.
For \Gb , its kinematic distance of $\sim$6.3 kpc \citep{Molinari:96,Qiu:12}
agrees with the parallax distance of $6.3^{+ 0.5}_{- 0.4}$ kpc.
}

\section{Galactic Distribution}

\begin{figure*}[htbp]
	\center
	\includegraphics[scale=0.7]{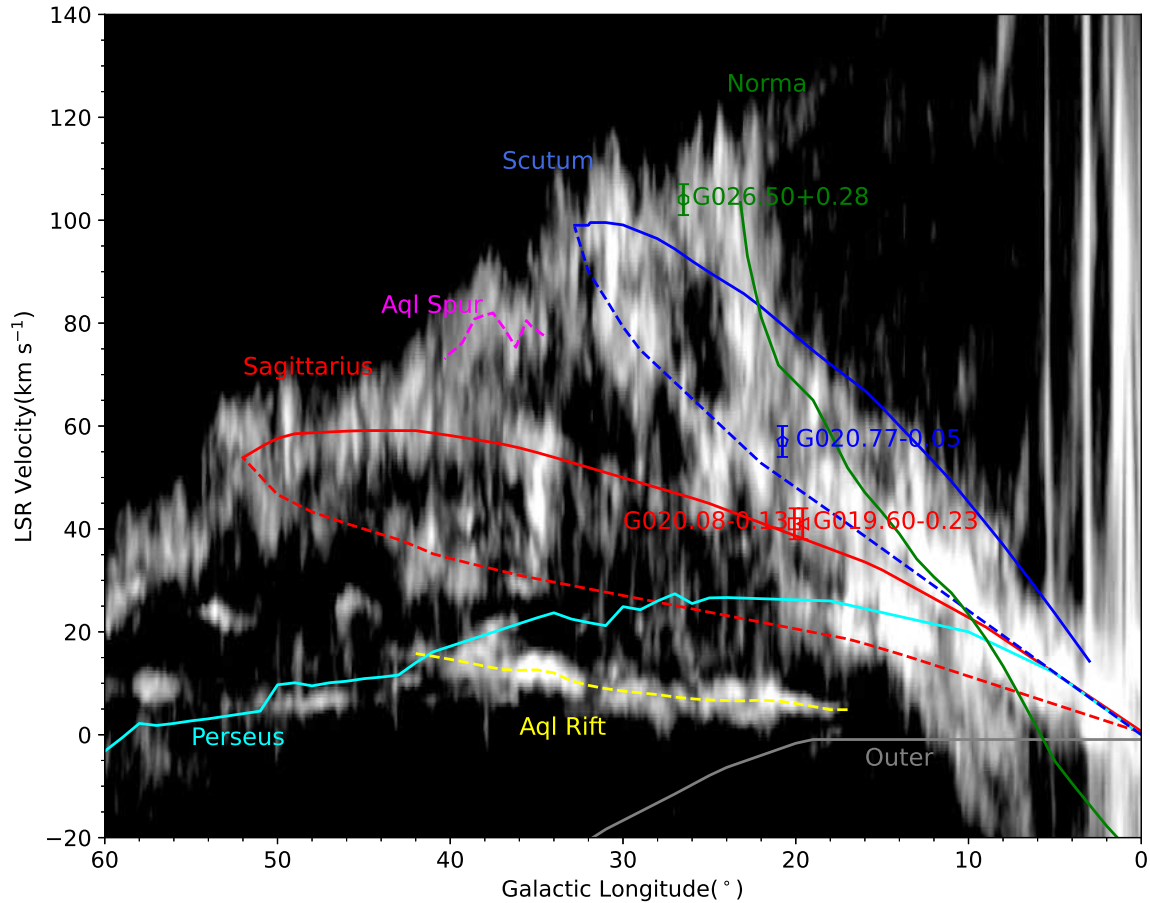}
	\caption{Location of masers superposed on a CO $(l,v)$ diagram
	  from \citep{Dame:01}; the spiral arm traces are from \citet{Reid:16}. Solid and dashed lines correspond to the
          near and far portions of arms interior to the Solar circle.
          The CO $(l,v)$ diagram is optimized to best display the intense inner-Galaxy arms, so the Perseus and Outer arms are not well represented here.   \label{lv}}
\end{figure*}

\begin{figure*}[htbp]
	\center
	\includegraphics[scale=0.56,trim=0 1.5cm 0 1.5cm,clip]{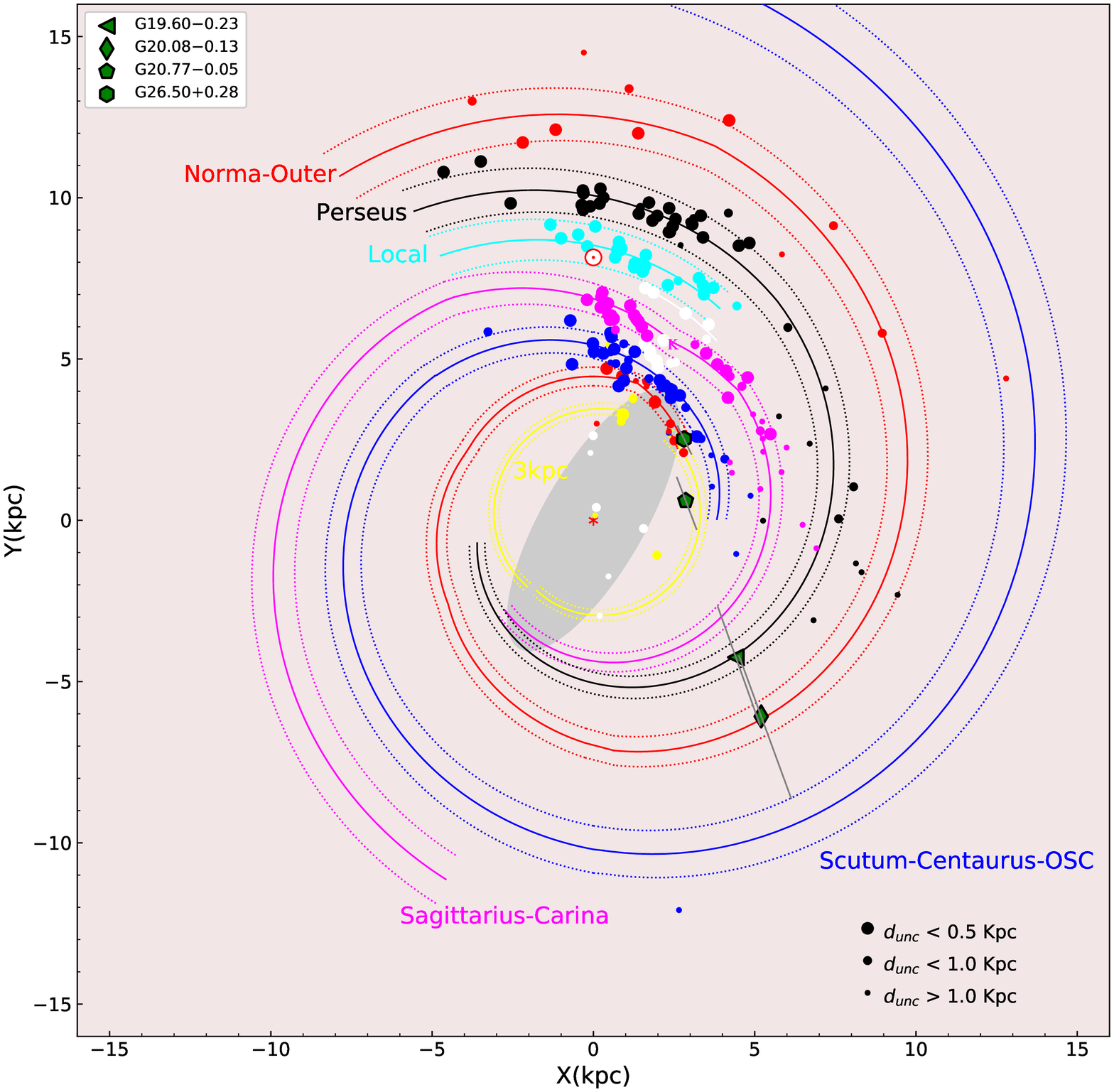}
	\caption{Plan view of the Milky Way from the north Galactic pole after \citet{Reid:19}.
		The four sources reported in this paper are indicated with green polygons
		(see legend in the upper left corner)\edit3{ with 1-sigma distance errors indicated}.
		Locations of maser sources with previously measured parallaxes are indicated with solid dots.
		\edit3{White dots indicate sources for which the arm assignments are unclear}.
		Solid curved lines trace the centers of spiral arm models from \citet{Reid:19},
		and the dotted lines enclose 90\% of sources.}  \label{xymap}
\end{figure*}

Since the major spiral arms of the Galaxy have long been known to
produce fairly well-defined arcs and loops in 21 cm and CO
longitude-velocity $(l,v)$ diagrams, the BeSSeL project assigns
maser sources to spiral arms based on their well-determined
positions in such diagrams --- even before their parallaxes
are measured. Figure~\ref{lv} shows traces of the spiral arms
in a CO $(l,v)$ diagram of the first Galactic quadrant, with
the positions of the present four maser sources overlaid.
Figure~\ref{xymap} shows the same four sources as green polygons
on the plane of the Galaxy along with the ~200 maser parallax
measurements (circles) from \citet{Reid:19}. The error bars
represent 1-sigma distance uncertainties.

Since the sources \Gd\ and \Ge\ have the same velocity and are separated
by less than one half degree on the sky, it seems likely that they are
physically associated. Judging from Figure~\ref{lv}, both appear to be
in the Sagittarius arm, although an association with the Perseus arm
cannot be ruled out. This is especially true since it is difficult to
determine the $(l,v)$ locus of the distant Perseus arm in this very
crowded section of the diagram \citep[see Figure 8 in][]{Reid:16}.
As shown in Figure~\ref{xymap}, our parallax distances are likewise
consistent with both sources being in the Sagittarius arm, although
judging from this figure alone the Perseus arm is favored.  Although
\Ge\ falls almost exactly on the Outer arm in Figure~\ref{xymap}, an
association with that arm can be confidently ruled out based on the
velocity separation of 43 \kms\ between this source and the Outer arm
locus in Figure~\ref{lv}. If both sources lie in the Sagittarius arm,
they hint at the need for a modification of the spiral arm fit
shown in Figure~\ref{xymap}, with the arm moving slightly farther out
and perhaps even merging with the Perseus arm at low Galactic longitudes.
However, tighter distance errors on these sources and others will be
required before such an adjustment can be confirmed.

\Gc\ presents an interesting case. Its $(l,v)$ position in Figure~\ref{lv}
places it in the near side of the Scutum arm, but its parallax distance
places it much closer to an extension of that arm to the far side
(Figure~\ref{xymap}).  These two results can be reconciled if the
source has a velocity anomalously low by about 20 \kms.  This is
entirely possible given the small Galactic radius of the source
and its proximity to the end of the Galactic bar. Indeed, as discussed
below, this source has very large peculiar motions in the plane of the
Galaxy. Based on these findings, we tentatively suggest that the far
portion of the Scutum arm extends from a Galactic longitude of 25\deg\
(corresponding to a Galactic azimuth of 90\deg) down to at least a
longitude of 20\deg\ (azimuth 100\deg). However, we caution that the
Norma arm could also be a candidate for extension, and the Norma and
Scutum arms may even merge there. Clearly more parallax measurements
are needed to untangle these possibilities.

\Gb\ is in a very crowded region of the inner Galaxy. Its $(l,v)$
position is consistent with both the near side of the Norma arm
and the far side of the Scutum arm. Our parallax distance of
$6.3^{+0.5}_{-0.4}$ kpc resolves this choice in favor of Norma.
The source is located near the end of the bar, where a large
number of high mass star-forming regions are known to exist.
Likely, the gravitational potential of the galactic bar
accumulates dust and gas there and intensifies the star-forming activity.

\begin{deluxetable}{lrrr}
\tablewidth{10pt} \tablecaption{Peculiar Motions \label{table:3d}}
\tablehead {
\colhead{Source}
 & \colhead{U$_{s}$} & \colhead{V$_{s}$} &
\colhead{W$_{s}$}
\\
\colhead{}     & \colhead{(\kms)}
 & \colhead{(\kms)} & \colhead{(\kms)}}
\startdata
\Gd &  $-$12 $\pm$    \phantom{0}8 &    7 $\pm$   63 &   $-$5 $\pm$   11\\
\Ge &  $-$13 $\pm$   \phantom{0}6 &   74 $\pm$   85 &   $-$8 $\pm$   11\\
\Gc &   31 $\pm$   16 &  $-$56 $\pm$   12 &    2 $\pm$   10\\
\Gb &   34 $\pm$   15 &  $-$14 $\pm$    \phantom{0}5 &   $-$8 $\pm$   10\\
\enddata
\tablecomments {The Galactic ``Univ'' model and solar motion found
by \citet{Reid:19} were used to calculate the peculiar motions.}
\end{deluxetable}

\edit3{
We calculate the peculiar (non-circular) motions for these four sources using the
methodology described in \citet{Reid:09} with the Galactic model and solar motion of \citet{Reid:19}.
The results are listed in Table~\ref{table:3d}.
For \Gd\ and \Ge, the peculiar motions are consistent with $U_s$ and $W_s$ magnitudes $\lax10$ \kms; the uncertainties in the $V_s$ values are too large for meaningful analysis.  These large $V_s$ uncertainties can be traced to a large
motion in Galactic longitude scaled by their Galactocentric radii.  These radii have $\pm2$ kpc
uncertainties, as their parallax distance uncertainties map almost directly to radii.
This is discussed in detail in \nameref{sec:appb}.}

Both \Gc\ and \Gb\ appear to have some
very large peculiar motions in the plane of the Galaxy.  They have motions
toward the Galactic center ($U_s$) of about 30 \kms, and these are significant
at the $2\sigma$ level.  Formally, \Gc\ has a peculiar component of 56 \kms\
counter to Galactic rotation.  However, this estimate is dependent on extrapolation
of the ``Universal'' rotation curve fitted to sources with Galactic radii $>4$ kpc.
Either \Gc\ is orbiting the Galaxy considerably slower than indicated by the
rotation curve, indicating a significantly elliptical orbit, and/or the rotation
curve is \edit3{a} too simplistic model for this region.

\section{Summary}
In order to extend our knowledge of the distances to sources
in the distant regions of the Milky Way, we measured the parallaxes
and proper motions of four masers associated with massive young stars.
\edit3{The results of \Gd\ and \Ge\ hint at a modification of the Sagittarius arm model of \citet{Reid:19}.} Near the Galactic
bar, many sources (\Gc, \Gb\ and others in \citet{Immer:19}) have significant
non-circular motions, suggesting the bar's strong gravitational
potential has a significant effect.

\facility{VLBA}

\acknowledgments
{This work was supported by the National Natural
Science Foundation of
China (Grant Numbers: 11933011, 11873019 and 11673066),
the Key Laboratory for Radio Astronomy,
Project of Xinjiang Uygur
Autonomous Region of China for Flexibly Fetching in Upscale Talents,
and Heaven Lake Hundred-Talent Program of Xinjiang
Uygur Autonomous Region of China.}


\section{APPENDIX A}

\label{sec:app} 
\begin{figure*}[htbp]
	\begin{minipage}[c]{0.22\linewidth}
	\includegraphics[scale=0.4,trim=0 4cm 0 6cm,clip]{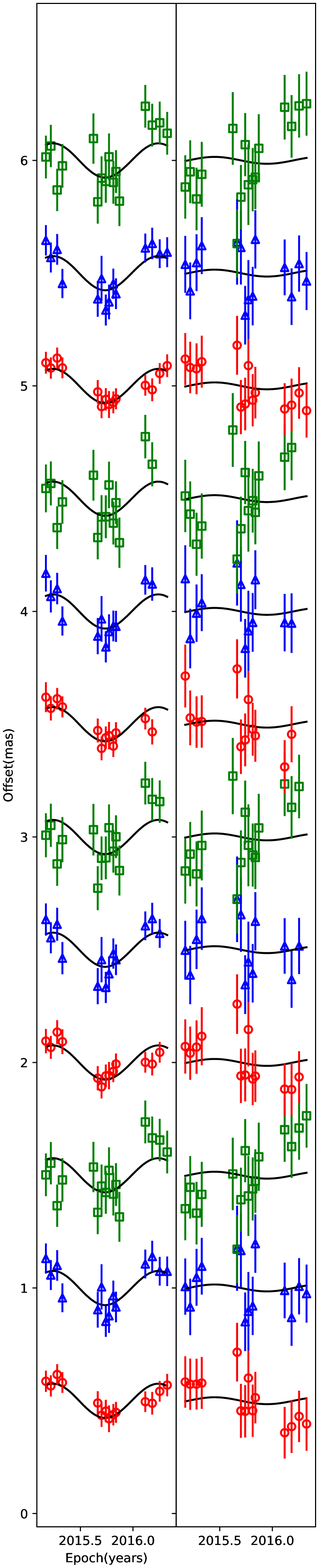}
	\centerline{\Gd}
	\centerline{\\}
\end{minipage}
\hfill
\begin{minipage}[c]{0.22\linewidth}
	\includegraphics[scale=0.4,trim=0 4cm 0 6cm,clip]{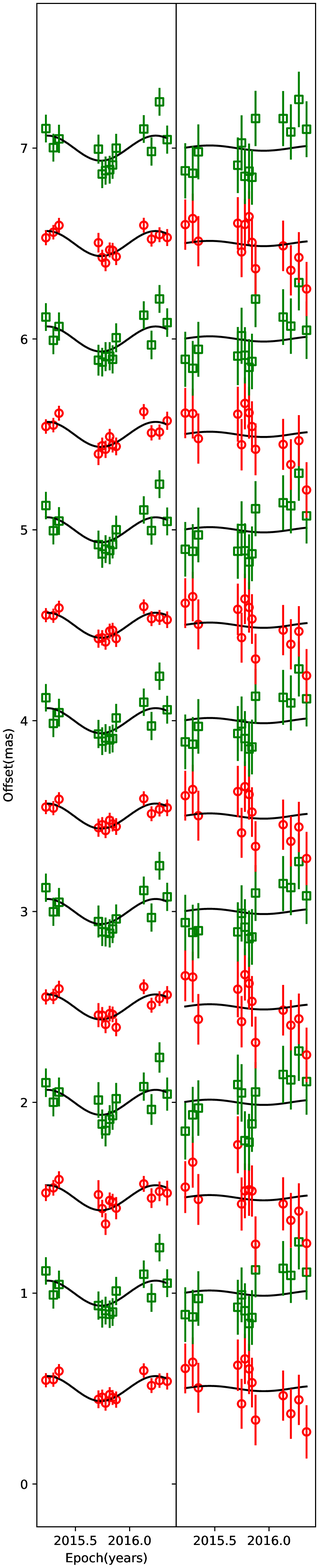}
	\centerline{\Ge}
	\centerline{\\}
\end{minipage}
\hfill
\begin{minipage}[c]{0.22\linewidth}
	\includegraphics[scale=0.4,trim=0 4cm 0 6cm,clip]{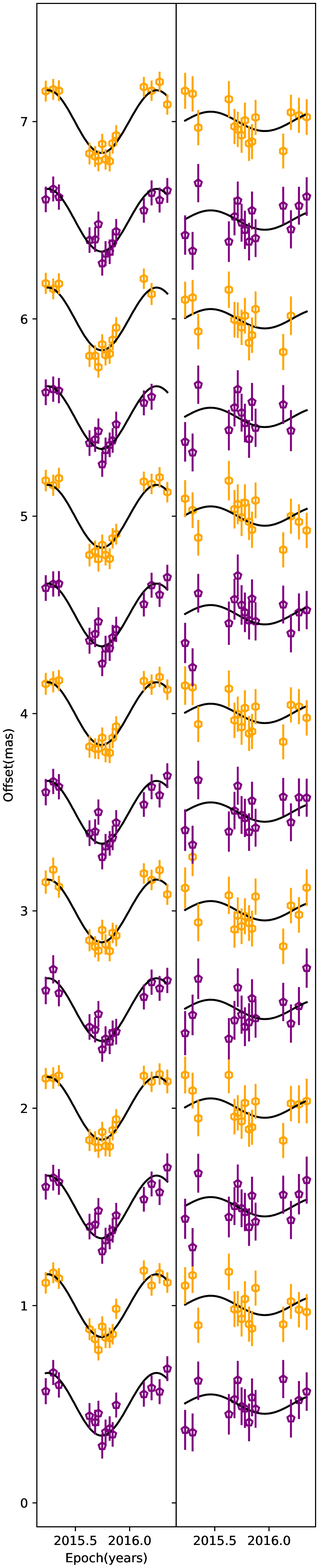}
	\centerline{\Gb}
	\centerline{\\}
\end{minipage}
\hfill
\begin{minipage}[c]{0.22\linewidth}
	\includegraphics[scale=0.4,trim= 0 1.5cm  0 2.5cm,clip]{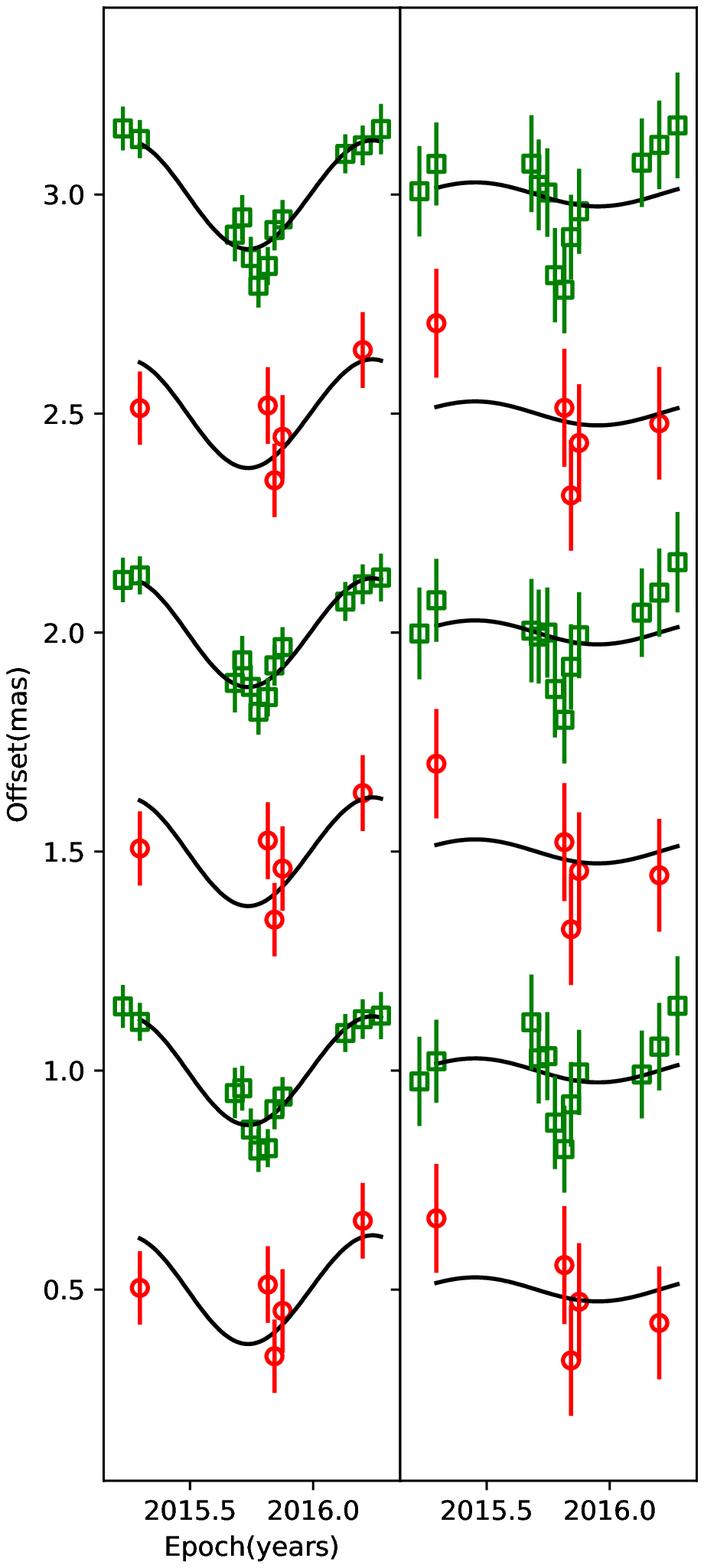}
	\centerline{\Gc}
\end{minipage}
	\caption{Results for combined parallax fit of \Gd, \Ge, \Gc, \Gb.
Left and right panels are combined parallax fit curves (black) in
R.A. and Dec directions, respectively. Different marks indicate different
background sources: red circles (J1821-1224), blue triangles (J1835-1248),
green squares (J1835-1115), purple pentagons (J1833-0323), and orange
hexagons (J1846-0651). \label{parallaxfit}}
\end{figure*}

Here, we show the details of the epochs observed (Table~\ref{table:observations}), observational parameters (Table~\ref{table:positions})
and parallax fits of the
individual sources (Table~\ref{table:detail} and Figure \ref{parallaxfit}).
The uncertainties of parallaxes and proper motions given in
Table~\ref{table:detail} are the formal fitting uncertainties.

\begin{deluxetable*}{lccccc}[htbp]
\tablecolumns{6} \tablewidth{0pc}
\tablecaption{Details of the Epochs Observed \label{table:observations}}
\tablehead{
\colhead{Epoch} & \colhead{Source}  & \colhead{Source} & \colhead{Source}
& \colhead{Source} }
\startdata
~ & G019.60$-$0.23 & G020.08$-$0.13 & G020.77$-$0.05 & G026.50+0.28  \\
E1  & 15 Mar 05  & & &                                 \\
E2  & 15 Mar 22 & 15 Mar 25 & 15 Mar 25  & 15 Mar 25   \\
E3  & 15 Apr 13 & 15 Apr 19 & 15 Apr 19  & 15 Apr 19   \\
E4  & 15 May 01 & 15 May 08 & 15 May 08  & 15 May 08   \\
E5  & 15 Aug 18 & 15 Aug 19 & 15 Aug 19  & 15 Aug 19    \\
E6  & 15 Sep 01 & 15 Sep 07 & 15 Sep 07  & 15 Sep 07   \\
E7  & 15 Sep 14 & 15 Sep 18 & 15 Sep 18  & 15 Sep 18   \\
E8  & 15 Sep 29 & 15 Oct 01 & 15 Oct 01  & 15 Oct 01   \\
E9  & 15 Oct 10 & 15 Oct 12 & 15 Oct 12  & 15 Oct 12   \\
E10 & 15 Oct 25 & 15 Oct 26 & 15 Oct 26  & 15 Oct 26   \\
E11 & 15 Nov 03 & 15 Nov 05 & 15 Nov 05  & 15 Nov 05   \\
E12 & 15 Nov 15 & 15 Nov 17 & 15 Nov 17  & 15 Nov 17   \\
E13 & 16 Feb 13 & 16 Feb 18 & 16 Feb 18  & 16 Feb 18   \\
E14 & 16 Mar 08 & 16 Mar 15 & 16 Mar 15  & 16 Mar 15   \\
E15 & 16 Apr 03 & 16 Apr 11 & 16 Apr 11  & 16 Apr 11   \\
E16 & 16 Apr 29 & 16 May 07 & 16 May 07  & 16 May 07   \\
\enddata
\end{deluxetable*}

\begin{deluxetable*}{lllcrcr}
\tablecaption{Positions and Brightnesses \label{table:positions}}
\tablehead{
\colhead{Source} & \colhead{R.A. (J2000)} &
\colhead{Dec. (J2000)} & \colhead{$\phi$} & \colhead{Brightness}
& \colhead{\Vlsr} & \colhead{NW beam} \\
\colhead{} & \colhead{$\mathrm{(^h\;\;\;^m\;\;\;^s)}$}
& \colhead{$(\degr\;\;\;\arcmin\;\;\;\arcsec)$} & \colhead{($^{\circ}$)}
& \colhead{(Jy/beam)} & \colhead{(\kms)}&
\colhead{(mas, mas, deg)} } \startdata
G019.60$-$0.23 & 18~27~38.0737 & $-$11~56~37.558             &
&  23.1 & 24.84 & 2.0 $\times$ 1.1 @ ~18      \\
J1821$-$1224   &18~21~23.27806 & $-$12~24~12.9282            & 1.6
& 0.023 &       & 2.3 $\times$   1.1 @    ~16      \\
J1835$-$1248   & 18~35~58.07853 & $-$12~48~51.5492           & 2.2
& 0.013 &       & 2.2 $\times$   1.2 @    ~15      \\
J1835$-$1115   & 18~35~19.57583 & $-$11~15~59.3116           & 2.0
& 0.026 &       & 2.1 $\times$   1.2 @    ~15      \\
G020.08$-$0.13 & 18~28~10.2867 & $-$11~28~47.892             &
& 7.8   & 46.43   & 1.7 $\times$ 0.4 @ $-$18     \\
J1821$-$1224   & 18~21~23.27806 & $-$12~24~12.9284           & 1.9
& 0.013 &       &  1.7 $\times$   0.4 @   $-$18    \\
J1835$-$1115   &  18~35~19.57583 & $-$11~15~59.3116           & 1.8
& 0.015 &       & 1.8 $\times$   0.4 @   $-$19      \\
G020.77$-$0.05 & 18~29~11.5819 & $-$10~49~58.566               &
& 1.9   & 58.60 & 1.9 $\times$ 1.0  @ ~15     \\
J1821$-$1224   & 18~21~23.27806 & $-$12~24~12.9287            & 2.5
& 0.007 &       &  2.0 $\times$   1.0 @    ~14     \\
J1835$-$1115   & 18~35~19.57583 & $-$11~15~59.3116           & 1.6
& 0.017 &       & 1.9 $\times$   1.0 @    ~16      \\
G026.50$+$0.28 &   18~38~40.7518 & $-$05~35~05.108           &
& 14.8   & 104.11 & 2.4 $\times$ 1.1 @ ~38       \\
J1846$-$0651   & 18~46~06.30026  & $-$06~51~27.7462       & 2.2
& 0.031  &       &  2.4 $\times$   1.0 @    ~37    \\
J1833$-$0323   & 18~33~23.90488 & $-$3~23~31.4442       & 2.6  &
0.035 &       &  2.3 $\times$   1.0 @    ~40   \\
\enddata
\tablecomments {$\phi$ is the angular separation between a maser
and its calibrator. The maser's absolute position, peak brightness,
size and position angle (PA), measured from North to East, of the naturally
weighted (NW) beam are listed for the first/second epoch for each
source. \Vlsr is the velocity of the phase-reference channel.}
\end{deluxetable*}
\clearpage

{\startlongtable
\begin{deluxetable*}{lcrccrrr}
\tabletypesize{\scriptsize}
\tablecolumns{8} \tablewidth{0pc} \tablecaption{Detailed Results
of Parallaxes and Proper Motions of Masers. \label{table:detail} }
	\tablehead{\\[-8pt] \colhead{Background} & \colhead{\Vlsr} & \colhead{Detected}
& \colhead{Parallax} & \colhead{$\mu_x$ }& \colhead{$\mu_y$}& \colhead{$\Delta
x$}& \colhead{$\Delta y$}\\
		\colhead{Source} & \colhead{(\kms)} & \colhead{epochs} & \colhead{(mas)}
& \colhead{(\masy)} & \colhead{(\masy)} & \colhead{(mas)}& \colhead{(mas)}}
	\startdata
	\multicolumn{8}{c}{\bf G019.60$-$0.23}\\\hline
	J1821$-$1224 & 24.84 & 1111 01111110 1111 &     0.059$\pm$0.008
&      $-$2.99$\pm$ 0.02 &      $-$6.62$\pm$ 0.04 &       0 &      0 \\
	& 38.75 & 1111 01111110 1111 &     0.071$\pm$0.009 &      $-$3.09$\pm$
0.02 &      $-$6.49$\pm$ 0.04 &     $-$679.513$\pm$      0.012 &
835.549$\pm$      0.017 \\
	& 42.20 & 1111 01111110 1100 &     0.061$\pm$0.010 &      $-$3.35$\pm$
0.03 &      $-$6.87$\pm$ 0.10 &     $-$986.537$\pm$      0.046 &
869.388$\pm$      0.078 \\
	& 43.82 & 1111 01111110 1110 &     0.058$\pm$0.010 &      $-$3.28$\pm$
0.03 &      $-$6.20$\pm$ 0.06 &     $-$1121.819$\pm$      0.025 &
1113.290$\pm$      0.037 \\
	\multicolumn{3}{c}{Combined fit} & 0.062$\pm$0.009  &   &  \\
	\hline
	J1835$-$1248 & 24.84 & 1111 01111110 1111 &     0.092$\pm$0.017
&      $-$2.89$\pm$ 0.04 &      $-$6.49$\pm$ 0.08 &       0 &      0 \\
	& 38.75 & 1111 01111110 1111 &     0.103$\pm$0.017 &      $-$3.00$\pm$
0.04 &      $-$6.35$\pm$ 0.07 &     $-$679.513$\pm$      0.012 &
835.549$\pm$      0.017 \\
	& 42.20 & 1111 01111110 1100 &     0.107$\pm$0.021 &      $-$3.18$\pm$
0.06 &      $-$6.72$\pm$ 0.10 &     $-$986.537$\pm$      0.046 &
869.388$\pm$      0.078 \\
	& 43.82 & 1111 01111110 1110 &     0.098$\pm$0.019 &      $-$3.17$\pm$
0.05 &      $-$6.05$\pm$ 0.09 &     $-$1121.819$\pm$      0.025 &
1113.290$\pm$      0.037 \\
	\multicolumn{3}{c}{Combined fit} & 0.100$\pm$0.018  &   &  \\
	\hline
	J1835$-$1115& 24.84 & 1111 11111111 1111 &     0.077$\pm$0.023 &
$-$2.75$\pm$ 0.06 &      $-$6.11$\pm$ 0.07 &       0 &      0 \\
	& 38.75 & 1111 11111111 1111 &     0.085$\pm$0.025 &      $-$2.86$\pm$
0.06 &      $-$5.98$\pm$ 0.09 &     $-$679.513$\pm$      0.012 &
835.549$\pm$      0.017 \\
	& 42.20 & 1111 11111111 1100 &     0.093$\pm$0.034 &      $-$3.05$\pm$
0.10 &      $-$6.40$\pm$ 0.13 &     $-$986.537$\pm$      0.046 &
869.388$\pm$      0.078 \\
	& 43.82 & 1111 11111111 1110 &     0.090$\pm$0.024 &      $-$3.01$\pm$
0.07 &      $-$5.72$\pm$ 0.10 &     $-$1121.819$\pm$      0.025 &
1113.290$\pm$      0.037 \\
	\multicolumn{3}{c}{Combined fit} &0.086$\pm$0.025  &   &  \\
	\hline
	Combined fit 2 &&&0.076$\pm$0.011&\\
	Average &&&&  $-$3.11$\pm$  0.03& $-$6.36$\pm$  0.06\\
	\hline
	\multicolumn{8}{c}{\bf G020.08$-$0.13}\\\hline
	J1821$-$1224
	& 47.29 & 111 00111111 1111 &     0.049$\pm$0.013 &      $-$3.24$\pm$
0.03 &      $-$6.34$\pm$ 0.10 &     0.698$\pm$    0.019 &    27.929$\pm$
0.038 \\
	& 46.43 & 111 00111111 1111 &     0.057$\pm$0.009 &      $-$3.37$\pm$
0.02 &      $-$6.79$\pm$ 0.08 &    0 &     0 \\
	& 42.44 & 111 00111111 1111 &     0.061$\pm$0.014 &      $-$3.04$\pm$
0.03 &      $-$6.69$\pm$ 0.08 &   210.409$\pm$    0.007 &    30.584$\pm$
0.018 \\
	& 41.58 & 111 00111111 1111 &     0.059$\pm$0.012 &      $-$3.11$\pm$
0.03 &      $-$6.67$\pm$ 0.07 &   208.070$\pm$    0.009 &    31.209$\pm$
0.021 \\
	& 40.50 & 111 00111111 1111 &     0.054$\pm$0.009 &      $-$3.06$\pm$
0.02 &      $-$6.67$\pm$ 0.08 &   210.415$\pm$    0.002 &    29.216$\pm$
0.004 \\
	& 39.85 & 111 00111111 1111 &     0.062$\pm$0.009 &      $-$3.03$\pm$
0.02 &      $-$6.86$\pm$ 0.08 &  $-$415.336$\pm$    0.002 &  $-$157.077$\pm$
0.005 \\
	& 37.05 & 111 00111111 1111 &     0.063$\pm$0.012 &      $-$3.22$\pm$
0.03 &      $-$6.59$\pm$ 0.09 &     6.160$\pm$    0.013 &  $-$274.918$\pm$
0.028 \\
	\multicolumn{3}{c}{Combined fit} & 0.058$\pm$0.011  &   &  \\
	\hline
	J1835$-$1115  & 47.29 & 111 00111111 1111 &     0.084$\pm$0.024
&     $-$3.19$\pm$ 0.06 &     $-$5.87$\pm$ 0.10 &     0.698$\pm$    0.019
&    27.929$\pm$    0.038 \\
	& 46.43 & 111 00111111 1111 &     0.089$\pm$0.022 &     $-$3.31$\pm$
0.06 &     $-$6.30$\pm$ 0.08 &    0 &    0 \\
	& 42.44 & 111 00111111 1111 &     0.095$\pm$0.021 &     $-$2.98$\pm$
0.05 &     $-$6.21$\pm$ 0.09 &   210.409$\pm$    0.007 &    30.584$\pm$
0.018 \\
	& 41.58 & 111 00111111 1111 &     0.089$\pm$0.023 &     $-$3.05$\pm$
0.06 &     $-$6.19$\pm$ 0.08 &   208.070$\pm$    0.009 &    31.209$\pm$
0.021 \\
	& 40.50 & 111 00111111 1111 &     0.087$\pm$0.022 &     $-$3.00$\pm$
0.06 &     $-$6.18$\pm$ 0.08 &   210.415$\pm$    0.002 &    29.216$\pm$
0.004 \\
	& 39.85 & 111 00111111 1111 &     0.095$\pm$0.021 &     $-$2.97$\pm$
0.05 &     $-$6.38$\pm$ 0.08 &  $-$415.336$\pm$    0.002 &  $-$157.077$\pm$
0.005 \\
	& 37.05 & 111 00111111 1111 &     0.097$\pm$0.022 &     $-$3.16$\pm$
0.06 &     $-$6.11$\pm$ 0.07 &     6.160$\pm$    0.013 &  $-$274.918$\pm$
0.028 \\
	\multicolumn{3}{c}{Combined fit} & 0.091$\pm$0.021  &   &  \\
	\hline
	Combined fit 2 &&&0.066$\pm$0.010&\\
	Average &&&&   $-$3.14$\pm$  0.02&  $-$6.44$\pm$  0.08\\
	\hline
	\multicolumn{8}{c}{\bf G020.77$-$0.05}\\\hline
	J1821$-$1224
	& 58.60 & 010 00000111 0100 &     0.094$\pm$0.047 &      $-$3.37$\pm$
0.14 &      $-$6.57$\pm$ 0.17 &      0 &     0 \\
	& 60.32 & 010 00000111 0100 &     0.093$\pm$0.049 &      $-$2.75$\pm$
0.14 &      $-$7.10$\pm$ 0.21 &     358.791$\pm$      0.010 &    $-$203.425$\pm$
0.017 \\
	& 62.05 & 010 00000111 0100 &     0.085$\pm$0.052 &      $-$3.25$\pm$
0.15 &      $-$6.82$\pm$ 0.19 &     193.117$\pm$      0.014 &    $-$112.847$\pm$
0.023 \\
	\multicolumn{3}{c}{Combined fit} & 0.091$\pm$0.043  &   &  \\
	\hline
	J1835$-$1115
	& 58.60 & 110 01111111 1110 &     0.127$\pm$0.015 &      $-$3.55$\pm$
0.04 &      $-$6.24$\pm$ 0.08 &      0 &      0 \\
	& 60.32 & 110 01111111 1110 &     0.135$\pm$0.014 &      $-$2.91$\pm$
0.04 &      $-$6.74$\pm$ 0.11 &     358.791$\pm$      0.010 &    $-$203.425$\pm$
0.017 \\
	& 62.05 & 110 01111111 1110 &     0.123$\pm$0.011 &      $-$3.40$\pm$
0.03 &      $-$6.45$\pm$ 0.09 &     193.117$\pm$      0.014 &    $-$112.847$\pm$
0.023 \\
	\multicolumn{3}{c}{Combined fit} & 0.129$\pm$0.013  &   &  \\
	\hline
	Combined fit 2 &&&0.124$\pm$0.013&\\
	Average &&&&  $-$3.27$\pm$  0.04& $-$6.55$\pm$  0.08\\
	\hline
	\multicolumn{8}{c}{\bf G026.50$+$0.28}\\\hline
	J1833$-$0323
	& 96.88 & 111 11111111 1111 &     0.136$\pm$0.015 &      $-$2.99$\pm$
0.04 &      $-$6.05$\pm$ 0.08 &     $-$462.539$\pm$      0.008 &
1161.731$\pm$      0.019 \\
	& 101.84 & 111 11111111 1111 &     0.135$\pm$0.018 &      $-$2.30$\pm$
0.05 &      $-$5.91$\pm$ 0.08 &     $-$51.324$\pm$      0.018 &    955.927$\pm$
0.041 \\
	& 102.49 & 111 11111111 1100 &     0.138$\pm$0.015 &      $-$2.33$\pm$
0.05 &      $-$5.64$\pm$ 0.11 &     149.468$\pm$      0.010 &    989.235$\pm$
0.022 \\
	& 102.92 & 111 11111111 1111 &     0.152$\pm$0.016 &      $-$2.37$\pm$
0.04 &      $-$5.79$\pm$ 0.09 &     $-$38.638$\pm$      0.013 &    936.896$\pm$
0.030 \\
	& 104.76 & 111 11111111 1111 &     0.137$\pm$0.015 &      $-$2.52$\pm$
0.04 &      $-$5.55$\pm$ 0.08 &     129.780$\pm$      0.028 &    1010.760$\pm$
0.056 \\
	& 106.59 & 111 11111111 1111 &     0.117$\pm$0.017 &      $-$2.76$\pm$
0.05 &      $-$6.87$\pm$ 0.07 &     607.880$\pm$      0.014 &    693.811$\pm$
0.029 \\
	& 107.13 & 111 11111111 1111 &     0.130$\pm$0.017 &      $-$2.36$\pm$
0.05 &      $-$5.81$\pm$ 0.08 &     160.330$\pm$      0.012 &    973.750$\pm$
0.029 \\
	\multicolumn{3}{c}{Combined fit} & 0.135$\pm$0.016  &   &  \\
	\hline
	J1846$-$0651
	& 96.88 & 111 11111111 1111 &     0.175$\pm$0.011 &      $-$2.98$\pm$
0.03 &      $-$6.23$\pm$ 0.06 &     $-$462.539$\pm$      0.008 &
1161.731$\pm$      0.019 \\
	& 101.84 & 111 11111111 1111 &     0.175$\pm$0.010 &      $-$2.29$\pm$
0.03 &      $-$6.08$\pm$ 0.06 &     $-$51.324$\pm$      0.018 &    955.927$\pm$
0.041 \\
	& 102.49 & 111 11111111 1100 &     0.186$\pm$0.016 &      $-$2.27$\pm$
0.05 &      $-$5.85$\pm$ 0.09 &     149.468$\pm$      0.010 &    989.235$\pm$
0.022 \\
	& 102.92 & 111 11111111 1111 &     0.189$\pm$0.011 &      $-$2.36$\pm$
0.03 &      $-$5.98$\pm$ 0.07 &     $-$38.638$\pm$      0.013 &    936.896$\pm$
0.030 \\
	& 104.76 & 111 11111111 1111 &     0.177$\pm$0.013 &      $-$2.50$\pm$
0.04 &      $-$5.72$\pm$ 0.08 &     129.780$\pm$      0.028 &    1010.760$\pm$
0.056 \\
	& 106.59 & 111 11111111 1111 &     0.157$\pm$0.013 &      $-$2.74$\pm$
0.04 &      $-$7.05$\pm$ 0.07 &     607.880$\pm$      0.014 &    693.811$\pm$
0.029 \\
	& 107.13 & 111 11111111 1111 &     0.172$\pm$0.010 &      $-$2.34$\pm$
0.03 &      $-$5.99$\pm$ 0.07 &     160.330$\pm$      0.012 &    973.750$\pm$
0.029 \\
	\multicolumn{3}{c}{Combined fit} & 0.176$\pm$0.012  &   &  \\
	\hline
	Combined fit 2 &&&0.159$\pm$0.012&\\
	Average &&&&   $-$2.51$\pm$  0.03&  $-$6.04$\pm$  0.06\\
	\enddata
	\tablecomments{In combined fit 2, we put all the data in the fitting
program and combine the individual error floors with the JMFIT values
in the data files.}
\end{deluxetable*}
}

\clearpage

\section{APPENDIX B}

\label{sec:appb}

%
%
%
%
%
%

Here we combine the steps documented in \citet{Reid:09} to obtain source
peculiar motions: \Us\ (toward the Galactic center), \Vs\ (in the direction of
Galactic rotation), \Ws\ (toward the north Galactic pole).  
\begin{equation*}
\begin{split}
U_s
=&~v_{Helio} \cos{b} \cos{(\ell + \beta)}\\
& - D \mu_{b} \sin{b} \cos{(\ell + \beta)}\\
& - D \mu_{\ell} \cos{b} \sin{(\ell + \beta)} \\
& - ({V_\odot} + {\Theta_0})\sin{\beta} \\
& + {U_\odot} \cos{\beta},
\end{split}
\end{equation*}
\begin{equation*}
\begin{split}
V_s
=&~v_{Helio} \cos{b} \sin{(\ell + \beta)}\\
& - D \mu_{b} \sin{b} \sin{(\ell + \beta)}\\
& + D \mu_{\ell} \cos{b} \cos{(\ell + \beta)} \\
& + ({V_\odot} + {\Theta_0})\cos{\beta} \\
& + {U_\odot} \sin{\beta}\\
& - \Theta_s,
\end{split}
\end{equation*}
\begin{equation*}
\Ws = (D \mu_{b} \cos{b} + \vhelio \sin{b} + W_\odot),
\end{equation*}
where \vhelio\ is the heliocentric radial velocity. 
$\mu_{\ell}$ and $\mu_{b}$ are proper motions in
Galactic longitude ($\ell$) and latitude ($b$) (converted from
those measured in the eastward and northward directions);
$\beta$ is Galactocentric azimuth (defined as 0 toward the Sun and
increasing in the direction of Galactic rotation);
$(U_\odot, V_\odot+\To, W_\odot)$ describe the non-circular motion of the Sun;
$\Theta_s$ is the rotation speed of Galaxy at source, and $D$ is its distance from Sun.

Using the following relations
$${{\sin{(\ell + \beta)}} \over \Ro } =  { \sin{\beta} \over {D_p} } ,~~$$
$$  R_s = - D_p \cos{(\ell + \beta)} + R_0 \cos{\beta},~~$$
where $D_p = D \cos{b}$, $R_0$ and $R_s$ are Galactocentric radii of Sun and sources,
we obtain $D \mu_{\ell} \cos{b} \sin{(\ell + \beta)} = \mu_{\ell} R_0 \sin{\beta}, D \mu_{\ell} \cos{b} \cos{(\ell + \beta)} = \mu_{\ell} (R_0 \cos{\beta} - R_s). $ Then,
\begin{equation}
\begin{split}
U_s
=&~v_{Helio} \cos{b} \cos{(\ell + \beta)}\\
& - D \mu_{b} \sin{b} \cos{(\ell + \beta)}\\
& - (\mu_{\ell} R_0 + {V_\odot} + {\Theta_0}) \sin{\beta}   \\
& + {U_\odot} \cos{\beta}, \label{Us}
\end{split}
\end{equation}
\begin{equation}
\begin{split}
V_s
=&~v_{Helio} \cos{b} \sin{(\ell + \beta)}\\
& - D \mu_{b} \sin{b} \sin{(\ell + \beta)}\\
& + (\mu_{\ell} R_0 + {V_\odot} + {\Theta_0})  \cos{\beta} -  \mu_{\ell} R_s\\
& + {U_\odot} \sin{\beta}\\
& - \Theta_s. \label{Vs}
\end{split}
\end{equation}

For \Gd\ and \Ge, uncertainties in the measured motions and distances lead to terms of $\sim10$ \kms\ for \Us\ in Eq. 1.
However, for \Vs\ the term $\mu_\ell R_s$ can contribute to a substantial uncertainty of
$$\sigma^2 = R^2_s \sigma^2_{\mu_\ell} +  \mu_\ell^2 \sigma^2_{R_s}~~,$$
dominated by the second term containing $\sigma^2_{R_s}$.
For $\mu_\ell \approx 7.2$ \masy\ and $\sigma_{R_s} \approx 2$ kpc,
this leads to an uncertainty of $\sigma \approx 70$ \kms.

\end{document}